\documentclass[12pt]{article}
\usepackage{epsf}
\usepackage{amsmath}
\usepackage{graphics}
\usepackage{cite}

\renewcommand{\thefootnote}{\#\arabic{footnote}}
\begin{document}

\newcommand{\gtrsim}{ \mathop{}_{\textstyle \sim}^{\textstyle >} }
\newcommand{\lesssim}{ \mathop{}_{\textstyle \sim}^{\textstyle <} }

\newcommand{\rem}[1]{{\bf #1}}

\renewcommand{\thefootnote}{\fnsymbol{footnote}}
\setcounter{footnote}{0}
\begin{titlepage}

\def\thefootnote{\fnsymbol{footnote}}

\begin{center}
\hfill astro-ph/yymmnnn\\
\hfill May 2007\\
\vskip .5in
\bigskip
\bigskip
{\Large \bf Cyclic Universe and Infinite Past}

\vskip .45in

{\bf Paul H. Frampton}

{\em Department of Physics and Astronomy,}

{\em University of North Carolina at Chapel Hill, NC 27599-3255, USA}

\end{center}

\vskip .4in
\begin{abstract}
We address two questions about the past for infinitely cyclic
cosmology. The first is whether it 
can contain an infinite length null geodesic into the past
in view of the Borde-Guth-Vilenkin (BGV) "no-go" theorem, The
second is whether, given that a small fraction of spawned
universes fail to cycle, there is an adequate probability for 
a successful universe after an infinite time. 
We give positive answers to both questions then show
that in infinite cyclicity the total number of universes
has been infinite for an arbitrarily long time.

\end{abstract}
\end{titlepage}

\renewcommand{\thepage}{\arabic{page}}
\setcounter{page}{1}
\renewcommand{\thefootnote}{\#\arabic{footnote}}

\newpage

In a recent article \cite{BF} a model for an infinitely cyclic
universe was proposed which addresses the entropy problem\cite{Tolman}
by appealing to phantom dark energy and a deflation mechanism
which reduces the large entropy accrued during expansion to
zero\cite{BF2} which is the constant adiabatic value during
contraction.

In this way, the entropy at turnaround is all displaced to
a volume exterior to our universe and this enables a cyclic
scenario to be consistent provided that the dark energy
has an equation of state satisfying $w = p/\rho < -1$.
It does not matter how much below $-1$ it is, although
observationally (and even anthropically \cite{BF2}) it
is easier to distinguish from a cosmological constant
the bigger is $\phi = |w + 1|$.

Several legitimate concerns must be addressed before we accept
that this is the correct solution about entropy
adopted by Nature. A couple of them will be addressed 
in this Letter. But first let us enumerate
a few concerns that will not be addressed here but which one hopes
to discuss in the future. 

One issue is the accomplishment of a turnaround as described
in \cite{BF} where there is an abrupt disintegration into
a large number $1/f^3$ of causal patches, each of which
separately contracts into a new universe. The scale factor
changes from $a(t)$ for expansion to $\hat{a}(t) = f a(t)$
for contraction.  No less important is a better understanding
of the bounce. Both of these issues
involve the critical density term for which exisiting 
derivations are based on extra dimensions\cite{Binetruy,Randall}.
A technical treatment in four spacetime dimensions is desirable in
each case to reassure us that the model is consistent without
commitment to string theory or extra dimensions. We know of no fatal flaw.

In the present article, our more modest aim is to address the question
of the infinite past during which necessarily there were an infinite
number of turnarounds and bounces each uniformly spaced by the
constant time period $\tau$.

Theorists are comfortable with an infinite future as
occurs in the standard model with a cosmological constant.
In that case the universe expands exponentially forever,
and other galaxies recede from ours to become invisible.
Entropy gradually increases.

There seems to be less widespread acceptance
of an infinite past.
One reason is the old worry about entropy\cite{Tolman}
that it must increase and so at a finite time in the past
would fall to zero. This is avoided in \cite{BF}.
Another possible concern is provided by arguments
about null geodesics into the past and whether the spacetime
manifold can be past complete; this is alleviated in the present Letter.

It is true that the infinite past is less familiar than the
infinite future, and surely an infinite past requires a 
cyclic model. 
At each turnaround a small
fraction ($f$) of universes will fail to cycle 
and we confirm that\footnote{It was D. Reichart who raised and answered this question of success versus failure.}
there is a high likelihood that after infinite cycles we live
in a successful universe. 
There is no reason that an infinite past is less viable
than an infinite future, although as we shall show it does
require somewhat unfamiliar concepts such as the inevitability
of infinite cyclicity and that the total number of universes
has been infinite for an arbitrarily long time. 

\bigskip
\bigskip

\newpage

\noindent \underline{{\it Past null geodesics}}

\bigskip
\bigskip

There is a general argument about past completeness
of the spacetime manifold which we address first.

\bigskip

We begin with the no-go theorem of \cite{BGV} which we shall
adapt for application to the more general case in \cite{BF},
as the original no-go theorem applies to past inflation. We 
shall show how this no-go theorem is by-passed, as the assumptions
no longer apply.

\bigskip

\noindent The metric is of the form
\begin{equation}
ds^2 = dt^2 - a(t)^2 d{\bf x}^2
\end{equation}
In this metric for a null geodesic the affine parameter $\lambda$
follows the relation
\begin{equation}
d\lambda \propto a(t) dt
\label{propto}
\end{equation}
We normalize the affine parameter to the present time $t = t_0$
by choosing with $a_0 = a(t_0)$
\begin{equation}
d\lambda = \left[ \frac{a(t)}{a_0} \right] dt
\end{equation}
so that $d\lambda/dt = 1$ when $t=t_0$.

\bigskip

\noindent Following \cite{BGV}, We multiply Eq.(\ref{propto}) by
the Hubble parameter $H=\dot{a}/a$ where a dot denotes derivative with respect to $t$
but now we integrate from an initial time $t_n = t_0 - n\tau$ up to
$t=t_0$ to obtain with $a_n = a(t_0 - n \tau), \lambda_n = \lambda (t_0 - n\tau)$
\begin{equation}
\int_{\lambda_n}^{\lambda_0} H(\lambda) d\lambda
= \frac{1}{a_0} \int_{a_n}^{a_0} da = nC
\label{integral}
\end{equation}
where in the cyclic model we have denoted the finite integral
\begin{equation}
\frac{1}{n a_0} \int_{a_n}^{a_0} da = C
\end{equation}
by the constant $C$.

\bigskip

\noindent The left hand side of Eq.(\ref{integral}) can be written as the average
of the Hubble parameter
\begin{equation} 
H_{av} \equiv \frac{1}{(\lambda_0-\lambda_n)}\int_{t_n}^{t_0} H(\lambda) d\lambda 
\label{Constraint}
\end{equation}
over $n$ cycles. In particular, it is important that $H_{av}$ in 
Eq.(\ref{Constraint}) is independent of the
integer $n$ because of cyclicity.

\newpage

\noindent Given Eq.(\ref{Constraint}), we find from Eq.(\ref{integral}) that
\begin{equation}
H_{av} = lim_{n \rightarrow \infty} \left[ \frac{nC}{(\lambda_0 - \lambda_n)}
\right]
\label{nullgeod}
\end{equation}
so that for $n \rightarrow \infty$, we find a backwards
null geodesic $(\lambda_0 - \lambda_n) \propto n$
of infinite length and the argument of \cite{BGV} does not apply.
Such a geodesic is exemplified by a photon propagating always
at the origin ${\bf x} = 0$ of the spatial coordinates.

Whether or not the past incompleteness arguments apply to the competing cyclic model
of \cite{ST1,ST2,ST3}, we take no position. In \cite{BGV}, it is argued that they
do, but the authors disagree\cite{PS}, so that jury is still out. But we do
assert that they do not apply to the model in \cite{BF}, as can be
seen directly from our Eq. (\ref{nullgeod}), where $(\lambda_0 - \lambda_n)$
necessarily becomes an infinite length past null geodesic
for $n \rightarrow \infty$, given the finiteness of both $H_{av}$
and $C$.

\bigskip
\bigskip

\noindent \underline{{\it Successful and failed universes}}

\bigskip
\bigskip

\noindent Now we turn to another issue. At each turnaround, a very large number $N$
of new universes is spawned. Let the number of universes at time $t=t_0-n\tau$ be
$\Sigma_n$. Then the total number now is $\Sigma_0 = N^n \Sigma_n$.

This is not quite right because although almost every causal patch contains
no photons and no matter, a tiny fraction $f<<1$ will contain one
or more photons and hence because of
pair production will fail to cycle and bounce prematurely.
Similarly any other matter such as a quark or lepton in the causal
patch will cause failure.
According to \cite{BF2}, this number is very small, generally $f < 1/N$
but we need to
examine the failed universes to assess the probability that we may live
in a successful rather than a failed universe now, after an infinite $n \rightarrow \infty$
of cycles.

\bigskip

Let us ignore any new universes spawned by failed universes. The
number of successful universes is given after $n$ cycles by
\begin{eqnarray}
\Sigma_0^{(successful)} & = & lim_{n \rightarrow \infty} [\Sigma_n (N-fN)^n] \cr
& = &  lim_{n \rightarrow \infty} \Sigma_n [(1-f)N]^n
\label{successful}
\end{eqnarray}

\bigskip

The number of failed universes, on the other hand, is
\begin{eqnarray}
\Sigma_0^{(failed)} & = & lim_{n \rightarrow \infty} \Sigma_n [fN + fN(1-f)fN + fN[(1-f)N]^2 + .....
+ fN[(1-f)N]^{(n-1)}]   \nonumber \\
& = & lim_{n \rightarrow \infty} \Sigma_n fN [(1-f)N]^{(n-1)} [1 - \{1/(1-f)N\}]^{-1} \nonumber \\
& = & lim_{n \rightarrow \infty} \Sigma_n fN [(1-f)N]^{n} [(1-f)N -1]
\label{failed}
\end{eqnarray}

\noindent The probability for a successful universe at present is
given by the ratio of Eq.(\ref{successful}) with the sum
of Eq.(\ref{successful}) and Eq.(\ref{failed}) which gives
\begin{equation}
P^{(successful)} = \frac{ [ (1-f)N - 1]}{(N-1)}
\label{probability}
\end{equation}
which for $N \gg 1$ and $f \ll 1$ is approximately $P^{successful} = (1 -f)$
similarly to each single turnaround, as expected.

\bigskip

This is non-trivial when both subsets are infinite and if it had been that failed
universes dominate instead, the model would have been untenable because our
universe would be infinitely unlikely. Fortunately, this is not the case.

\bigskip
\bigskip

\noindent \underline{{\it Total number of universes}}

\bigskip
\bigskip

Last but certainly not least, we study the total number of universes
versus time in the past.

\bigskip

Suppose that $\Sigma_n < \infty$ for some finite $n$. Then going back another $n^{'}$
cycles we have $\Sigma_{n+n^{'}} = \Sigma_n N^{-n^{'}}$. N satisfies $N > 1$ (actually
$N >> 1$) so for some $n^{'}$ the integral part of $\Sigma_{n+n^{'}} = 1$ and cyclicity fails.
Therefore no finite $\Sigma_n$ is permitted for any finite $n$. In particular, the present
number of universes must be $\Sigma_0 = \infty$, as expected after an infinite number
of cycles.

\bigskip

\noindent More subtle is the value of
\begin{eqnarray}
\Sigma_{\infty} & = & lim_{n \rightarrow \infty} (\Sigma_0 N^{-n}) \nonumber \\ 
& = & \Sigma_0 [lim_{n \rightarrow \infty} (N^{-n})]
\label{Sigmainfty}
\end{eqnarray}
which is indeterminate as the product of infinity ($\Sigma_0$) times zero.
This requires some recourse to cardinality and the transfinite numbers
of set theory\cite{Cantor,Halmos}, depending on the level of rigor
demanded.

\bigskip

In set theory the lowest transfinite is $\aleph_0$ (Aleph-zero) and the
simplest assumption is that the number of universes is always $\aleph_0$,
the cardinality of the primes, the integers or the rational numbers.
When $\aleph_0$ is multiplied by a finite number $N$, it remains $\aleph_0$.
This holds for any finite $n$ in Eq.(\ref{Sigmainfty}) and so extends back
an arbitrarily long time in the past. 
For the infinite past, one cannot really say anything from
Eq.(\ref{Sigmainfty}).

\bigskip

\noindent At each turaround the number of universes increase by a gigantic
factor $N$ but $N \times \infty = \infty$ so in that sense the total number
remains infinite. 

\bigskip

\noindent The process is not time-reversal invariant and the global entropy of
all universes increases with time consistent with the second law of thermodynamics.
Considering only our universe, however, the entropy as well as the
density and temperature are cyclic and never infinite. This is as
near to infinite cyclicity as seems possible consistent
with statistical laws. The old problem confronting Tolman\cite{Tolman}
is avoided by removing entropy to an unobservable exterior region;
one may say in hindsight 
that the problem lay in considering only one universe. 

\bigskip

\noindent \underline{Summary}

\bigskip
\bigskip

We have argued that an infinite past time is consistent
and that the cyclic model of \cite{BF} is an exemplar.
The presence of patches which fail to cycle is not a problem as
after an infinite number of cycles the probability
of being in a successful universe as we find ourselves
is practically one. Also, it is mandatory that the total number
of universes is infinite for an arbitrarily long
time into the past. We assume the total number
of universes is constant and equal to $\aleph_0$ (Aleph-zero).
This idea is unfamiliar but
appears to us to be an inevitable concomitant of an infinite past.

\bigskip
\bigskip
\bigskip
\bigskip

\begin{center}

{\bf Acknowledgements}

\end{center}

\bigskip
This work was supported in part by the
U.S. Department of Energy under Grant No. DE-FG02-06ER41418.

\end{document}